# Random Matrix Theory and the Evolution of Business Cycle Synchronisation, 1886 – 2006

Paul Ormerod

Volterra Consulting, London, UK and Institute of Advance Study, University of Durham, UK

March 2008

pormerod@volterra.co.uk

*Abstract*

*The standard literature on business cycle convergence relies upon the estimation of an empirical correlation matrix of time series data of macroeconomic aggregates in the various countries.*

*The major study by Bordo and Helbing (2003) analyses the business cycle in Western economies over the 1881-2001 period. They examine four distinct periods in economic history and conclude that there is a secular trend towards greater synchronisation for much of the $20^{th}$ century, and that it takes place across these different regimes.*

*However due to the finite size of both the number of economies and the number of observations, a reliable determination of the correlation matrix may prove to be problematic. The structure of the correlation matrix may be dominated by noise rather than by true information. Random matrix theory was developed in physics to overcome this problem, and to enable true information in a matrix to be distinguished from noise.*

*Using a very similar data set to Bordo and Helbing, I use random matrix theory, and the associated technique of agglomerative hierarchical clustering, to examine the evolution of convergence of the business cycle between the capitalist economies over the long-run.*

*Contrary to the findings of Bordo and Helbing, it does not seem possible to speak of a 'secular trend' towards greater synchronisation over the period as a whole. During the pre-First World War period the international business cycle does not exist in any meaningful sense. The cross-country correlations of annual real GDP growth are indistinguishable from those which could be generated by a purely random matrix. The periods 1920-1938 and 1948-1972 do show a certain degree of synchronisation – very similar in both periods in fact – but it is very weak. In particular, the cycles of the major economies cannot be said to be synchronised during these periods. Such synchronisation as exists in the overall data set is due to meaningful co-movements in sub-groups.*

*So the degree of synchronisation has evolved fitfully, and it is only in the most recent period, 1973-2006, that we can speak of a strong level of synchronisation of business cycles between countries.*

1.   **Introduction**

Bordo and Helbing (2003) examine the evolution of the synchronisation of the business cycle in 16 capitalist economies over the 1880 to 2001 period. They use data that covers four distinct eras with different international monetary regimes. The four eras are 1880-1913 when much of the world adhered to the classical Gold Standard, the interwar period (1920-1938), the Bretton Woods regime of fixed but adjustable exchange rates (1948-1972), and the modern period of managed floating among the major currency areas (1973 to 2001).

The authors conclude that 'using three different methodologies that there is a secular trend towards increased synchronization for much of the twentieth century and that it occurs across diverse exchange rate regimes'.

These methodologies rely on empirical estimates of the correlation matrix of time series data of macroeconomic aggregates in the various countries. However due to the finite size of both the number of economies and the number of observations, a reliable determination of the correlation matrix may prove to be problematic. The structure of the correlation matrix may be dominated by noise rather than by true information. In other words, the apparent increase in sychronisation might be due to noise in the correlation matrix rather then to genuine differences in information. If this is the case, we cannot rely on apparent differences in values of correlation matrices calculated over different time periods.

Random matrix theory has been successfully applied by physicists to financial market data in order to overcome this problem (for example, Laloux et.al. (1999), Bouchaud and Potters (2000), Mantegna and Stanley (2000), Plerou et.al. (2000)). Ormerod and Mounfield (2002) apply the technique to recent quarterly real GDP growth data in the main EU economies.

This short paper investigates the application of the concepts of random matrix theory to the correlations between the annual growth rates of real GDP to a very similar set of economies over a very similar time period to that of Bordo and Helbing.

Section 2 discusses the data and methodology, and the results are set out in section 3.

## 2    Data and methodology

The annual real GDP data for 16 countries 1885-1994 is taken from Maddison (1995). The 1995-2006 data is from the IMF database. Strictly speaking, the two sources are not exactly comparable since the Maddison data is in real Geary-Khamis dollars and the IMF in domestic currency, but given that we are working with annual GDP growth, this is of little consequence.

The countries[1] are: Australia, Austria, Belgium, Canada, Denmark, Finland, France, Germany, Italy, Japan, Netherlands, New Zealand, Norway, Sweden, United Kingdom and United States.

Bordo and Helbing note that 'Output correlations have been the perhaps most frequently used measures of business cycle synchronization. According to this measure, national cycles are synchronized if they are positively and significantly correlated with each other. The higher are the positive correlations, the more synchronized are the cycles. Compared with concordance correlations, measuring synchronization with standard contemporaneous correlations is more stringent, as the latter require similarities in both the direction and magnitudes of output changes'.  The same approach is used here, namely the correlations between annual real GDP growth rates are examined.

The data during and immediately after the two world wars give rise to considerable distortions in the analysis.  For example, as a result of the massive bombing, both conventional and atomic, of Japan in 1945, output fell by 50 per cent. In Germany, output fell 29 per cent in 1945 and a further 41 per cent in 1946. The largest fall in a single year was in fact 59 per cent in Austria in 1945. Output in France dropped by 16 per cent in 1917 and a further 21 per cent in 1918. Given that the approach being used requires similarities not just in sign but also in the size of output changes, the years 1914-1919 and 1939-1947 are omitted from the analysis.

---

[1] In the Maddison data set, Swiss GDP data is available but only from 1900 on an annual basis.  However, using data 1900-2006 shows that the results are very robust to the inclusion or otherwise of Switzerland, so it is omitted from the main analysis because of the lack of Swiss growth rate data 1886-1900

The distribution of the eigenvalues of *any* random matrix has been obtained analytically (Mehta, 1991). In particular, the theoretical maximum and minimum values can be calculated. We compare the eigenvalues of the correlation matrix of the data series in which we are interested with the theoretical maximum and minimum values of those of a random matrix of similar dimension.

In order to assess the degree to which an empirical correlation matrix is noise dominated we can compare the eigenspectra properties of the empirical matrix with the theoretical eigenspectra properties of a random matrix. Undertaking this analysis will identify those eigenstates of the empirical matrix who contain genuine information content. The remaining eigenstates will be noise dominated and hence unstable over time.

For a scaled random matrix **X** of dimension N x T, (i.e where all the elements of the matrix are drawn at random and then the matrix is scaled so that each column has mean zero and variance one), then the distribution of the eigenvalues of the correlation matrix of **X** is known in the limit T, N $\to \infty$ with Q = T/N $\geq$ 1 fixed. The density of the eigenvalues of the correlation matrix, $\lambda$, is given by:

$$\rho(\lambda) = \frac{Q}{2\pi} \frac{\sqrt{(\lambda_{max} - \lambda)(\lambda - \lambda_{min})}}{\lambda} \qquad \text{for } \lambda \in [\lambda_{min}, \lambda_{max}] \qquad (1)$$

and zero otherwise, where $\lambda_{max} = \sigma^2 (1 + 1/\sqrt{Q})^2$ and $\lambda_{min} = \sigma^2 (1 - 1/\sqrt{Q})^2$ (in this case $\sigma^2 = 1$ by construction).

The eigenvalue distribution of the correlation matrices of matrices of actual data can be compared to this distribution and thus, in theory, if the distribution of eigenvalues of an empirically formed matrix differs from the above distribution, then that matrix will not have random elements. In other words, there will be structure present in the correlation matrix.

To analyse the structure of eigenvectors lying outside of the noisy sub-space band the Inverse Participation Ratio (IPR) may be calculated. The IPR is commonly utilised in to quantify the contribution of the different components of an eigenvector to the magnitude of that eigenvector (e.g. Plerou et. al. 1999).

Component $i$ of an eigenvector $v_i^\alpha$ corresponds to the contribution of time series $i$ to that eigenvector. That is to say, in this context, it corresponds to the contribution of economy $i$ to eigenvector $\alpha$. In order to quantify this we define the IPR for eigenvector $\alpha$ to be

$$I^\alpha = \sum_{i=1}^{N} (v_i^\alpha)^4$$

Hence an eigenvector with identical components $v_i^\alpha = 1/\sqrt{N}$ will have $I^\alpha = 1/N$ and an eigenvector with one non-zero component will have $I^\alpha = 1$. Therefore the inverse participation ratio is the reciprocal of the number of eigenvector components significantly different from zero (i.e. the number of economies contributing to that eigenvector).

## 3    Results

I first of all examine the period 1886-1913, very similar to the Gold Standard period of Bordo and Helbing. The largest eigenvalue of the correlation matrix has a value of 2.86 and the second largest 2.30.

Given the number of countries and number of observations, the theoretical upper limit of the eigenvalues of a purely random matrix is 3.08. However, (1) only holds in the limit, and so I examined the possible existence of small-sample bias. Computing the eigenvalues of the correlation matrix of 10,000 such random matrices[2] did in fact suggest a some small sample bias, with the highest value being 3.68. Only 234 out of the 10,000 largest eigenvalues were above the theoretical value of 3.08.

So hypothesis that the correlation matrix of annual real output growth over this period is entirely dominated by noise and contains no true information cannot be rejected. In other words, during the late 19$^{th}$ century and the years immediately prior to the First World War, there was no synchronisation at all of the business cycles of the capitalist economies.

A graphical representation of the issue is provided by the technique of agglomerative hierarchical clustering. (Kaufman and Rousseeuw (1990)). The approach constructs a hierarchy of clusters. At first, each observation is a small cluster by itself. Clusters are merged until only one large cluster remains which contains all the observations. At each stage the two 'nearest' clusters are combined to form one larger cluster. In the results presented here, the distance between two clusters is the average of the dissimilarities between the points in one cluster and the points in the other cluster[3].

Figure 1 plots the hierarchical clustering obtained from the correlation matrix of annual output growth 1886-1913.

---

[2] Which each column is a separately drawn random normal variable with mean 0 and standard deviation 1
[3] The analysis was carried out using the command 'agnes' in the statistical package S-Plus, with the default options of metric = 'euclidean' and method = 'average'.

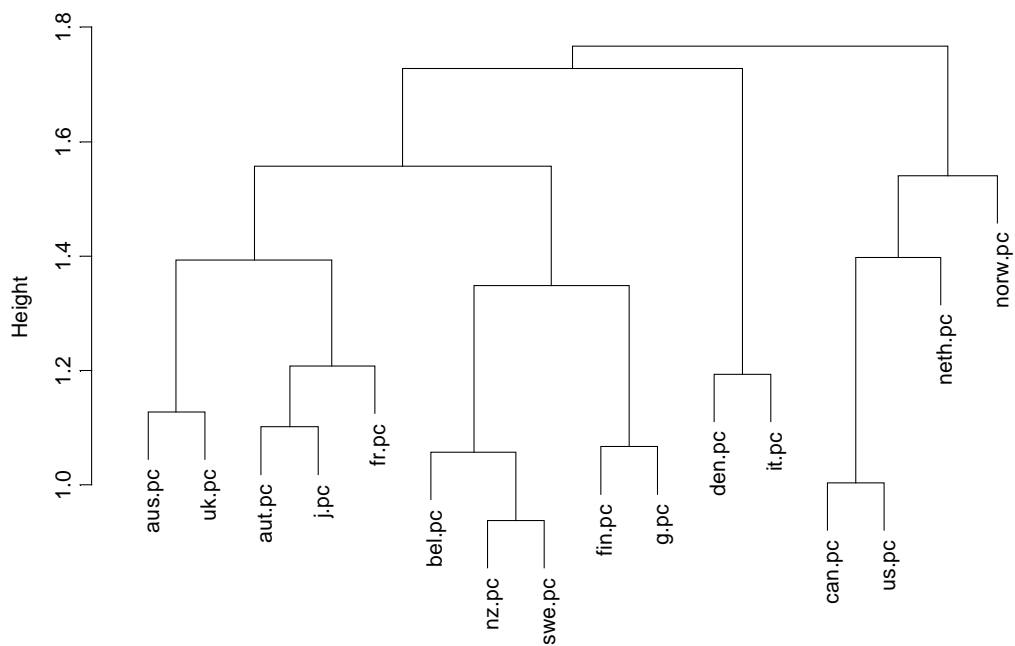

**Figure 1** *Agglomerative hierarchical clustering of the correlation matrix of annual real GDP growth rates in 16 countries, 1886-1913; the countries are in general obvious from their labels, though 'aus' is Australia and 'aut' is Austria. The suffix 'pc' is used to denote percentage change i.e. the correlation matrix of the percentage growth rates*

A certain amount of exposition of the chart may be useful. The horizontal axis is of no significance to the observed structure, and relevant information is on the vertical axis. The vertical axis measures the distance at which the economies are merged into clusters. So, rather bizarrely, the first two economies to be merged into a cluster, in other words the two whose synchronization of the business cycle was highest, are New Zealand and Sweden.

The random nature of the synchronization during this period is reflected in the fact that few of the clusters make any meaningful economic sense. The merging of Canada and the United States and the UK and Australia at an early stage appears sensible, but none of the others have any real economic rationale.

In contrast, the hierarchical clustering of the 1973-2006 data yields clusters which have a ready economic interpretation.

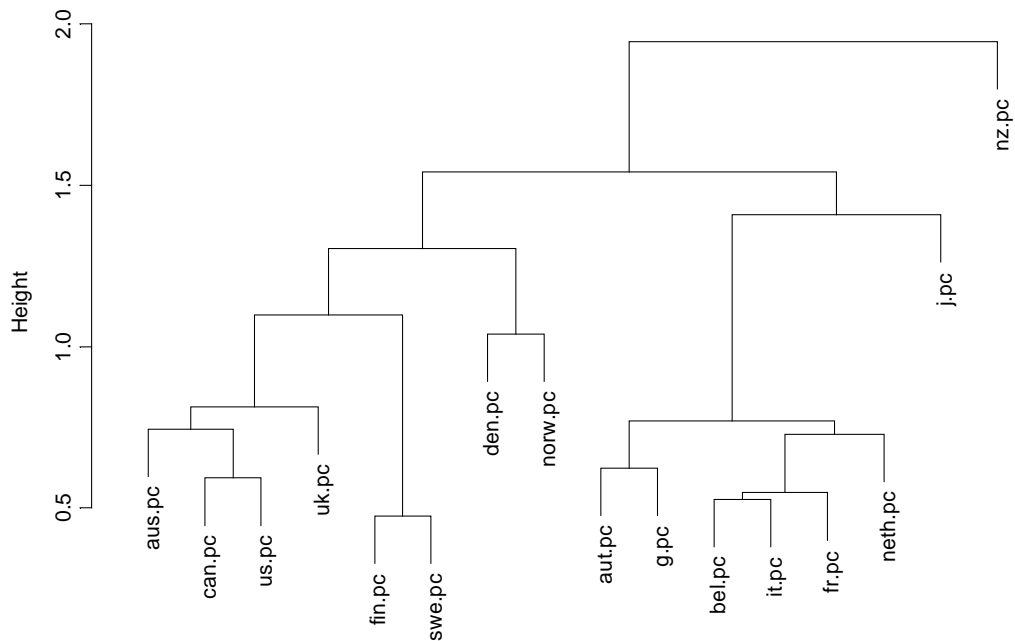

**Figure 2**  *Agglomerative hierarchical clustering of the correlation matrix of annual real GDP growth rates in 16 countries, 1973-2006*

Japan, which of course experienced a major asset deflation around 1990 and as a result a decade of poor growth, and New Zealand are rather isolated from the rest. But the main groupings are readily identifiable: the Anglo-American bloc of the US, UK, Canada and Australia; the main EU bloc of Austria and Germany, Belgium, Italy and France, and the Netherlands; a Scandinavian group of Finland and Sweden and Denmark and Norway.

The existence of true information in the correlations over this period is shown by the value of the principal eigenvalue of the correlation matrix, 6.76. This compares to the value given by (1) of 2.84, and the highest value of 3.35 obtained in 10,000 calculations of the eigenvalues of the correlation matrix of a random matrix of the same dimension, with only 217 being above 2.84. The second empirical eigenvalue is 2.60 and so within the random range.

The eigenvector associated with the principal eigenvalue mirrors the information displayed in Figure 2. The IPR is 13.51, compared with the maximum potential value of 16 when all 16 countries are contributing equally to the vector. The values for each economy in this vector are Australia 0.22, Austria 0.27, Belgium 0.29, Canada 0.29, Denmark 0.23, Finland 0.23, France 0.32, Germany 0.27, Italy 0.31, Japan 0.15, Netherlands 0.31, New Zealand 0.07, Norway 0.16, Sweden 0.23, UK 0.25, US 0.27. The value for New Zealand is distinctly different from all the others. The fact that most of the other individual elements are similar in size shows that this vector corresponds to a collective motion of all of the GDP growth time series. It is therefore a measure of the degree to which the growth of different countries is correlated.

So during the period prior to the First World War, it is not meaningful to speak of an international business cycle, but one definitely exists during the 1973-2006 period.

The inter-war period, 1920-1938, exhibits a certain amount of structure in terms of synchronisation, but less decisively so than the 1973-2006 period. The value of the main eigenvalue, 5.97, is considerably higher than the theoretical value from (1) of 3.68, but this period in particular has a shortage of observations, and the empirical upper limit obtained by 10,000 simulations of a random matrix is 4.36. Interestingly, the main economies of the period - US, UK, Germany, France and Italy – exhibit no meaningful synchronisation. The principal eigenvalue of the correlation matrix of these economies is 2.08 compared to the value given by (1) of 2.44 and the simulated highest value is 2.88. So such true synchronisation as exists is between small groups of countries. Belgium and France; Germany, Austria and Netherlands are the clearest examples, as well of course as the US and Canada.

The Bretton Woods period, 1948-1972, has, perhaps surprisingly, more in common with the inter-war period than the 1973-2006 one. The main eigenvalue is above the maximum given by (1), 4.65 compared to 3.24, and it is also above the maximum value of 3.86 obtained empirically by 10,000 simulations of a random matrix. However, the 6 major economies (adding Japan to the list) exhibit no difference from purely random correlations. The principal eigenvalue of the correlation matrix of these 6 economies is 2.10 compared to the random maximum of 2.39. The main country groupings which give some true synchronization to the full data set are somewhat different from the inter-war period: the US and Canada are the same, but otherwise there is a group of France, Germany and Austria and a 'Fringe Europe' one of the UK, Sweden and Finland, although Belgium is also in this group.

The evolution over time of the degree of synchronization can be examined. The trace of the correlation matrix is conserved, and is equal to the number of independent variables for which time series are analysed. For the correlation matrix of the main 6 economies[4], for example, the trace is equal to 6 (since there are 6 time series). The closer the 'market' eigenmode (i.e. eigenmode 1) is to this value the more information is contained within this mode i.e. the more correlated the movements of GDP. The market eigenmode corresponds to the largest eigenvalue, $\lambda_{max}$. The degree of information contained within this eigenmode, expressed as a proportion, is therefore $\lambda_{max}/N$.

To follow the evolution of the degree of business cycle convergence over time we may analyse how this quantity evolves temporally. The analysis is undertaken with a fixed window of data. Within this window the spectral properties of the correlation matrix formed from this data set are calculated. In particular the maximum eigenvalue is noted for each period.

Figure 3 plots the evolution of the principal eigenvalue of the correlation matrix for the main 6 economies over the 1948-2006 period, using a window of 12 years. More precisely, it sets out the evolution of $\lambda_{max}/N$, where N = 6. So the first observation is $\lambda_{max}/N$ for the 1948-1959 period, the second for the 1949-1960 period, and so on.

---

[4] These have consistently made up around 85 per cent of the total output of the 16 countries in the data set

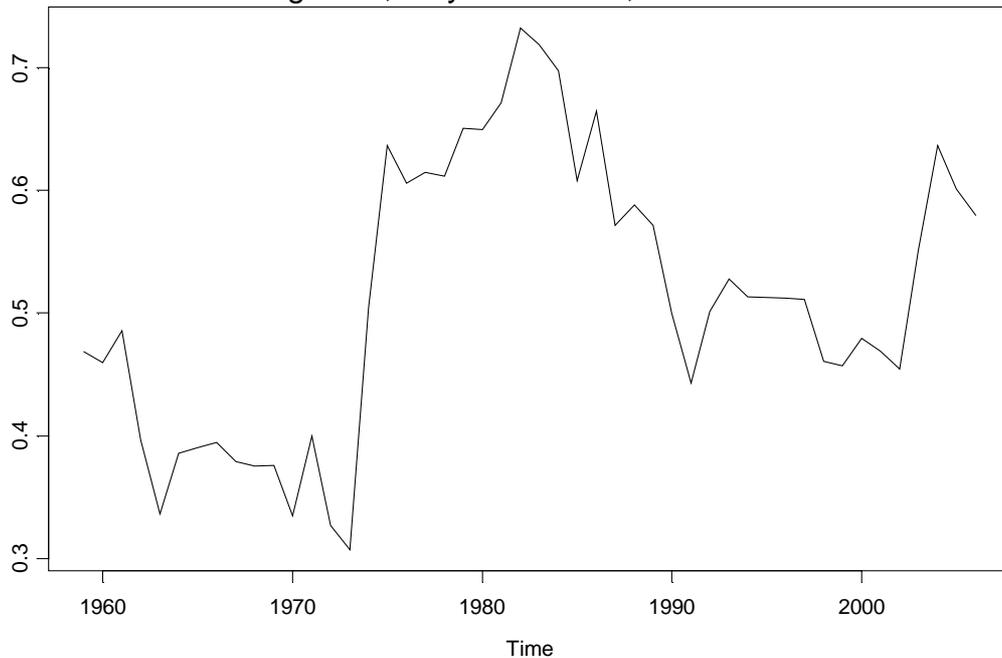

**Figure 3**   *The temporal evolution of the degree of information content in the maximum eigenvalue of the empirical correlation matrix formed from the time series of annual real GDP growth for the main world economies of the US, UK,, Germany, France, Italy and Japan, 12 year windows, 1948-2006*

Over the 1948-1959 period, for example, the first observation in the chart, the 'market' eigenvalue took up just under 50 per cent of the total of the eigenvalues, indicating a reasonable but not dramatic degree of convergence of their business cycles. But then, advancing year by year there is a distinct trend fall, until over the 1962-1973 period, a minimum is reached where the maximum eigenvalue is only 30 per cent of the total.

The common experience of the major shocks of the mid-1970s leads to a dramatic rise in the degree of convergence of their business cycles, reaching a peak in the period 1972-1983. This remained high for several years, before declining in the light of Japan's problems and German re-unification, which temporarily dislocated German convergence with the other main EU economies, for example (Ormerod and Mounfield, op.cit.). In more recent years, convergence has risen again in the relatively calm condition which have prevailed since the mid-1990s.

## 4  Discussion

There is a large literature on the degree of business cycle convergence amongst the main Western economies over the most recent decades. A key question is whether or not the cycles have become more synchronised. On this, the literature is essentially inconclusive.

Bordo and Helbing (op.cit) take a much longer perspective and examine the business cycle in Western economies over the 1881-2001 period. They examine four distinct periods in economic history and conclude that there is a secular trend towards greater synchronisation for much of the $20^{th}$ century, and that it takes place across these different regimes.

Most of the analytical techniques used in the business cycle convergence literature rely upon the estimation of an empirical correlation matrix of time series data of macroeconomic aggregates in the various countries. However due to the finite size of both the number of economies and the number of observations, a reliable determination of the correlation matrix may prove to be problematic. The structure of the correlation matrix may be dominated by noise rather than by true information.

Random matrix theory was developed in physics to overcome this problem, and to enable true information in a matrix to be distinguished from noise. It has been successfully applied in the analysis of financial data.

Using a very similar data set to Bordo and Helbing, I use random matrix theory, and the associated technique of agglomerative hierarchical clustering, to examine the evolution of convergence of the business cycle between the capitalist economies.

The results confirm that there is a very clear amount of synchronisation of the business cycle across countries during the 1973-2006 period. In contrast, during the pre-First World War period it is not possible to speak of an international business cycle in any meaningful sense. The cross-country correlations of annual real GDP growth are indistinguishable from those which could be generated by a purely random matrix.

However, in contrast to Bordo and Helbing, it does not seem possible to speak of a 'secular trend' towards greater synchronisation over the 1886-2006 period as a whole. The periods 1920-1938 and 1948-1972 do show a certain degree of synchronisation – very similar in both periods in fact – but it is weak. In particular, the cycles of the major economies cannot be said to be synchronised during these periods. Such synchronisation as exists in the overall data set is due to meaningful co-movements in sub-groups.

So the degree of synchronisation has evolved fitfully, and it is only in the most recent period, 1973-2006, that we can speak of a strong level of synchronisation of business cycles between countries.

More detailed analysis of the evolution of synchronisation of the 6 major economies (US, UK, Germany, France, Italy, Japan) in the post-Second World War period, suggests that it can vary considerably over relatively short periods of time. There is a distinct trend towards *less* synchronisation during the 1950s and 1960s, and it is during the period of the major shocks to the Western economies in the 1970s and early 1980s that synchronisation was at its peak, supporting the finding of Bordo and Helbing that common shocks are a major source of synchronisation.